\documentstyle[preprint,epsfig,aps]{revtex}

\epsfverbosetrue
\draft
\newcommand{\Si}[1]{\mbox{Si$_{#1}$}}

\begin{document}
\title{Growth Pattern of Silicon Clusters}
\author{Atul Bahel, Jun Pan,\footnote{Also, the Department of Physics,
New York University, New York, New York 10003-6621.}
and Mushti V. Ramakrishna}
\address{The Department of Chemistry, New York University,
New York, NY 10003-6621.}
\date{Mod. Phys. Lett. B {\bf 9}, XXXX (1995)}
\maketitle

\begin{abstract}

Tight-binding molecular dynamics simulated annealing technique is
employed to search for the ground state geometries of silicon clusters
containing 11-17 atoms.  These studies revealed that layer formation is
the dominant growth pattern in all these clusters.  Fullerene-like
precursor structures consisting of fused pentagon rings are also
observed.  The atoms in all these clusters exhibit pronounced
preference for residing on the surface.

\end{abstract}
\vspace{0.1 in}

Silicon is a semiconductor with diamond lattice structure.  Due to its unique
electrical properties it is the most important technological material
in electronics industry.  As the drive towards miniaturization of
electronic devices accelerates, there is a pressing need to
investigate the properties of low-dimensional semiconductors
\cite{Pool:90,Corcoran:90,Bjornholm:90,Siegel:93}.  For this
reason, the past few years have witnessed enormous scientific activity
in the experimental and theoretical investigations of silicon clusters,
slabs, and wires of nanometer dimensions
\cite{Siegel:93,Elkind:87,Jarrold:89,Phillips:88,Jelski:88,Kaxiras:89,Bolding:90,Patterson:90,Swift:91,Roth:94,Tomanek:86,Menon:91,Menon:93-1,Menon:94}.

Since naked silicon clusters are
highly reactive, they are mostly synthesized in a molecular beam under
high vacuum conditions \cite{Elkind:87,Jarrold:89}.  The number density
of available clusters is so low that diffraction based structural
investigations are not feasible under these experimental conditions.
Consequently, the experimentalists are attempting to infer the
structures of these clusters through indirect evidence derived from the
reactivities (or sticking coefficients) of these clusters with various
reagents.  In support of these experimental efforts we have recently
started to investigate the structures of silicon clusters using
an empirical tight-binding Hamiltonian.  The results of these
calculations are presented here.

Menon and Subbaswamy have constructed a non-orthogonal tight-binding
Hamiltonian for silicon clusters \cite{Menon:93-2,Ordejon:94}.  This
method is equivalent in spirit to the well known extended H\"{u}ckel
method \cite{Lowe:78}.  The computational cost of this method is
intermediate between the methods based on empirical potential
and {\em ab initio} molecular orbital
techniques.  In this method all the valence
electrons are explicitly included in the determination of the
electronic energy.  However, the interaction matrix elements between
non-orthogonal atomic orbitals are adjustable parameters.  Menon and
Subbaswamy have related these matrix elements to Harrison's universal
parameters appropriate for the description of bulk Si
\cite{Harrison:80}.  In addition, they added four new parameters
for the description of the silicon clusters.  These additional
parameters are derived by fitting to the \Si{2} bond length and
vibrational frequency and to the overall size-dependent cohesive energy
curve of clusters \cite{Menon:93-2,Ordejon:94}.  While the parameters
are fit to either the experimental or calculated data,
none of these parameters are fit to any
of the cluster structures.  Full details of the Hamiltonian and
computational methods are described elsewhere
\cite{Menon:93-2,Ordejon:94}.

As the size of the cluster grows, the number of structural isomers
increases exponentially, with the result that searching the complete
configuration space for the global potential energy minimum becomes a
formidable task.  However, by combining the tight-binding method with
the molecular dynamics \cite{Allen:87} simulated annealing technique we
can efficiently search the cluster configuration space and determine the
ground state geometry \cite{Bahel:95,Pan:95}.  We used this TB-MD
technique in all the calculations reported here.

Figure 1 displays the lowest energy structures obtained using this
TB-MD technique for \Si{N} clusters in the $N$ = 11-17 atom size
range.  Figure 2 displays alternate structures obtained for some of
these clusters.  The \Si{11} structure consists of a tetragon-pentagon
sandwich in the anti-prism geometry with face caps at the bottom and
top.  This structure may also be viewed as an icosahedron with one
missing atom.  We also obtained a related structure for \Si{11} whose
cohesive energy is only slightly higher than the global minimum
presented in Fig. 1.  This local minimum structure, displayed in Fig. 2,
consists of a tetragon-tetragon anti-prism sandwich with one face cap at the
bottom and two face caps at the top.  This structure is also one of two
possible candidate ground state structures of \Si{11} found by
Rohlfing and Raghavachari using {\em ab initio} molecular orbital
techniques \cite{Rohlfing:90}.

The structure of \Si{12} is similar to \Si{11}, consisting of a
pentagon-pentagon anti-prism sandwich with face caps at the top and
bottom \cite{Bahel:95,Pan:95}.  This structure is highly spherical and
each atom is five-fold coordinated.  Such a spherical cage structure
has not been predicted or observed for any other 12-atom elemental
cluster.  The compound clusters boranes and carboranes are the only
other clusters that form icosahedral cage structures.

Adding a face cap to \Si{12} gives the lowest energy \Si{13}
structure.  We also observed an isomer of this cluster, consisting of
4-4-4-1 layers (Fig. 2).  The difference in the cohesive energies of
these two structures is less than 0.01 eV/atom.  We also considered
several alternative structures for this cluster, but most of them are
either unstable or high energy local minima.  For example, placing a Si
atom inside the cage of \Si{12} yields a high energy local minimum for
\Si{13}.  Likewise, structures based on hexagon-hexagon sandwiches are
found to be unstable.

The \Si{14} structure is similar to one of the isomers of \Si{13}.  It
consists of a 4-4-4 layer structure and two adjacent face caps.  By
suitably rotating this structure, we may also describe it as a
pentagon-pentagon prism sandwich with two caps each at the top and the
bottom.  The pentagon prism is somewhat distorted and displaced.  We
also found an isomer of this cluster, consisting of a \Si{4} subunit
riding on top of a \Si{10} structure (Fig. 2).  The cohesive energies
of these two structures are within 0.02 eV/atom of each other.  This is
the first cluster that may be viewed as a super-cluster consisting of
stable subunits.  We also considered a bi-capped hexagonal anti-prism
as a candidate for the ground state of \Si{14}.  However, this
structure proved to be unstable, indicating that six-atom ring
structures are still not favored in these small clusters.  Like \Si{14}
cluster, \Si{15} assumes a layer structure consisting of 1-5-3-5-1
layers.  We did not observe any nearly degenerate isomers for \Si{15}.

All the clusters up to \Si{15} may be described as stacked layers
consisting of four- or five-membered rings.  These
layers are suitably terminated at the top  and bottom.  However,
\Si{16} is quite unlike any of these clusters.  It is an open cage
consisting of fused pentagons, reminescent of the small fullerenes.
Since the smallest fullerene must contain at least
twenty atoms \cite{Curl:91,Boo:92,Fowler:92}, it appears that \Si{16} is the
precursor to the formation
of such fullerene cages \cite{RK:94}.

We also found a fullerene-like cage structure for \Si{17}, but it is
not the global minimum.  Instead, a super-cluster consisting of stable
\Si{7}
and \Si{10} subunits is found to be the global minimum.
An alternative super-cluster consisting
of \Si{6} and \Si{11} subunits is found to be a local minimum (Fig.
2).  The structures
of these subunits are identical to those of the
corresponding isolated clusters
\cite{Rohlfing:90,Krishnan:85}.

These results indicate that formation of layers is the dominant
nucleation and growth
pattern in silicon clusters.  The exception to this rule is \Si{16},
whose structure is similar to that of small fullerenes consisting of
fused pentagon rings.  Some clusters seem to prefer to segregate into
smaller subunits consisting of 4, 6, 7, or 10 atoms.  Each of these
subunits is a magic number cluster \cite{Krishnan:85}, thus explaining
the ability of larger clusters to form super-structures.

In summary, we determined the structures of small silicon clusters
through TB-MD simulations.  These simulations revealed that \Si{11} is
an incomplete icosahedron, \Si{12} is a complete icosahedron, \Si{13}
is a surface capped icosahedron, \Si{14} is a 4-4-4 layer structure with
two adjacent face caps,
\Si{15} is a 1-5-3--5-1 layer structure, \Si{16} is a partially closed
cage consisting of fused pentagons reminescent of small fullerenes, and
\Si{17} is a super-structure consisting of \Si{7} + \Si{10} subunits.
The atoms in all these clusters exhibit a strong preference to lie on
the surface rather than inside.  Formation of stacked layers is the
primary mechanism of nucleation and growth in these clusters.

This is the seventh paper in this series on {\em Chemistry of
Semiconductor Clusters}.  This research is supported by the New York
University Research Challenge Fund and the Donors of The Petroleum
Research Fund (ACS-PRF \# 26488-G), administered by the American
Chemical Society.  The support of New York University Scientific
Visualization Center is gratefully acknowledged.  All the computations
reported in this paper were carried out on an IBM 580 Workstation.  The
graphics presented here are generated using the XMol program from the
Research Equipment Inc. and the University of Minnesota Supercomputer
Center.

\begin{figure}
\caption{The lowest energy structures of \Si{N} ($N$ = 11-17)
clusters determined using the tight-binding molecular dynamics
simulated annealing technique.}
\end{figure}

\begin{figure}
\caption{Alternate structures of \Si{N} ($N$ = 11, 13, 14, 17)
found in tight-binding molecular dynamics
simulations.  The ground state structures of these clusters are
displayed in Fig. 1.}
\end{figure}

FIGURES NOT SUPPLIED 

\end{document}